\documentclass{iau} 
\usepackage{graphicx}

\title[The IMF with ELTs] 
{The Initial Mass Function in the ELT era}

\author[Kieran Leschinski \& Jo\~ao Alves]   
{Kieran Leschinski$^1$ \and Jo\~ao Alves$^1$}

\affiliation{$^1$Department of Astrophysics, University of Vienna, \\ 
T\"urkenschanzstrasse 17, 1180 Vienna, Austria \\ email: {\tt kieran.leschinski@univie.ac.at} \\[\affilskip]
}

\pubyear{2018}
\volume{347}  
\jname{Early Science with ELTs}
\editors{Norbert Przybilla, Karen Pollard \& Annalisa Calamida}
\begin{document}

\maketitle

\begin{abstract}
The initial mass function (IMF) is an important, yet enigmatic aspect of the star formation process. The two major open questions regarding the IMF are: is the IMF constant regardless of environment? Is the IMF a universal property of star formation? The next generation of extremely large telescopes will allow us to observe further, fainter and more compact stellar clusters than is possible with current facilities. In these proceeding we present our study looking at just how much will these future observatories improve our knowledge of the IMF.

\end{abstract}

\firstsection 
\section{Introduction}


The stellar initial mass function (IMF) plays a significant role in astrophysical processes through out the universe. The consequences of the IMF are seen on almost all distance scales, from the stellar composition of young star clusters to the chemical enrichment of the intergalactic medium. Cosmological simulations also include various versions of the IMF. For all its importance though, it is still a poorly understood aspect of the star formation process and is described by an empirically derived function. Past and present studies of the IMF rely in being able to reliably determine the masses of stars in a young population. Such studies are therefore naturally restricted by the detection limits and resolution of modern telescopes. The current generation of telescopes have allowed us to determine the shape of the sub-solar IMF within the solar neighbourhood, and place limits on its slope in the solar- and super-solar mass range outside the Milky Way (see e.g. \cite[Bastian et al. 2010]{bastian10}, \cite[Da Rio et al.2009]{dario09}, \cite[Geha et al. 2013]{geha13}).

The future generation of extremely large 30-40\,m class telescopes will allow the astronomical community to observe star forming regions which are fainter and more distant than anything yet observed. The extremely large telescope (ELT) \cite[(Gilmozzi \& Spyromilio 2007)]{elt}, combined with the MICADO wide-field near infrared adaptive optics (AO) assisted camera \cite[(Davies et al. 2010)]{micado}, will enable observations at the diffraction limit of the 39\,m telescope (7 mas in J, 12 mas in K) and will be able to detect stars down to $\sim$29 mag in J-band (vega). 
In these proceedings we report on our efforts to determine just how useful MICADO and the ELT will be for future studies of the IMF.

\section{Simulations}

	
\begin{figure}
	\centering
	\includegraphics[width=\textwidth]{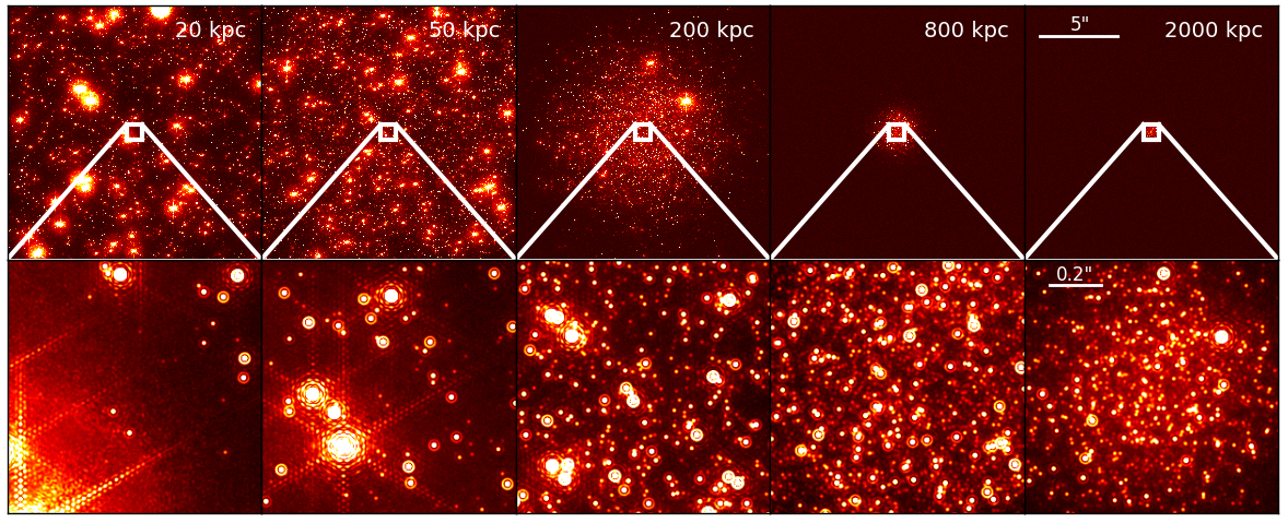} 
	\caption{Five of the 42 clusters used in this study. The 1000 M$_\odot$ clusters are placed at distances ranging from 20\,kpc out to 2 \,Mpc. The top row shows the field of view of the central MICADO detector ($\sim$16'', while the bottom row shows a 1'' cut-out from the centre of the detector.}
	\label{fig:clusters}
\end{figure}

The goal of our current study is to answer the question: What is the lowest reliably observable mass star in young clusters for a given distance and density that will be observable with the ELT and MICADO? To answer this we created models of 42 young stellar clusters, located at distances ranging from 8 kpc out to 5 Mpc and with surface densities of between 100 and 100\,000 stars arcsec $^{-2}$. To put this in context, the Orion nebula cloud (ONC) contains about 5000\,M$_\odot$ within a $\sim$1\,pc$^2$ region. If this were located in the Large Magellanic Cloud (LMC) the surface density would be $\sim$3000\,stars\,arcsec$^{-2}$. 

We used SimCADO  the instrument data simulation software for MICADO at the ELT, in order to generate mock observation of these clusters.
We then used an iterative PSF subtraction photometry technique to detect and extract the stars from the images. For a full description of this process see \cite[Leschinski et al (in prep)]{leschinski2018}. The extracted stars where cross-matched against the input catalogue and the instrumental flux compared with the star's intrinsic flux. The stars were finally sorted into mass bins to find the (sigma-clipped) standard deviation of the original vs. detected flux ratios. Our metric for defining lowest ``reliably observable'' mass was to take the upper mass limit of the bin where the standard deviation of this original vs. detected flux ratio was greater than 10\%. In other words, the mass bin where more than 32\% of stars had detected fluxes differing by more than 10\% from their original input flux.

\section{Results and Discussion}


Table \ref{tab:results} lists the lowest ``reliably observable'' masses for the cluster configurations which we studied. Two main effects come into play when determining this mass limit: sensitivity and resolution. For the more distant clusters the limit is defined by sensitivity, regardless of surface density. For example at a distance of 2\,Mpc only stars heavier than 2\,M$_\odot$ are detectable for all surface densities below 30\,000 stars arcsec$^{-2}$ (equivalently $\sim$300 stars pc$^{-2}$). Only above this density are there enough stars with M$>$2\,M$_\odot$ that crowding becomes significant. Conversely closer to Earth crowding becomes a problem at much lower surface densities, because many more stars are above the detection limit.
In the LMC stars around the hydrogen burning limit ($\sim$0.1\,M$_\odot$) will be detectable in regions with surface densities of less than $\sim$100 stars arcsec$^{-2}$ (equivalently~$\sim$1500 stars pc$^{-2}$). However as stellar density and crowding increases, so too does the effective detection limit.
At surface densities of 10\,000~stars~arcsec$^{-2}$ (equivalently $\sim$150\,000 stars pc$^{-2}$) the detection limit is $\sim$0.5\,M$_\odot$ - on par with many contemporary studies of much less dense regions in the LMC (e.g. \cite[Da Rio et al. 2009]{dario09}). For the sake of comparison, one of the most massive young clusters known in the Milky Way, Westerlund 1, contains~$\sim$150\,000~stars~pc$^{-2}$.

The most important results from table \ref{tab:results} can be summarised as follows:

\begin{itemize}
    \item \textbf{Brown dwarf mass function:} ($<$0.1\,M$_\odot$) will be accessible anywhere in the Milky Way for regions with densities lower than 1000 stars arcsec$^{-2}$. This will allow the structure of the brown dwarf ``knee'' (\cite[Kroupa 2001]{kroupa01}) to be thoroughly investigated and will help determine whether the shape of the IMF remains constant irrespective of environment.
    
    \item \textbf{Shape of Low mass IMF:} ($<$0.4\,M$_\odot$) and the position of the turn over will be possible in the cores of dense regions ($<$10\,000 stars arcsec$^{-2}$) in the nearest galaxies (LMC, SMC, etc). This will shed light on the question of whether the IMF is universal.
    
    \item \textbf{The IMF of cluster cores:} is currently inaccessible to current facilities. These are limited to observing the outskirts of young clusters for all but the nearest clusters due to crowding in the centre. We have shown that ELT observations will easily be able to cope with surface densities in excess of 1000 stars arcsec$^{-2}$, and will still provide good quality data for surface densities in excess of 5000 stars arcsec$^{-2}$.
    
    \item \textbf{Super-solar IMF:} will be observable in galaxies at distances out to 5\,Mpc. This will afford us a much better understanding of the extra-galactic IMF and provide many more data points to help answer the question of the universality of the IMF.
    
\end{itemize}

It should also be stated that while many studies have estimated the properties of the IMF in galaxies outside the local neighbourhood, these studies rely on integrated light and assumptions of the underlying stellar population. The only way to verify the accuracy of these assumptions and therefore the reliability of the results is to count the individual stars which make up the regions used in extra-galactic IMF studies. It should be stressed that this will be one of the  greatest advantage of using ELTs to study the IMF: the ability to resolve individual stars in galaxies not just within the galactic neighbourhood, but in galaxies in and beyond our local group. This will allow an important bridge to be be built between the techniques of determining the IMF from individual sources and from using population models and integrated light.

\begin{table}
	\centering
	\caption{Lowest reliably observable mass in the centre of a young stellar cluster for given star surface densities and distances with MICADO at the ELT.}
	\vspace{0.3cm}
	\label{tab:results}

	\begin{tabular}{lccccccc}
	\hline
	\hline
	Distance	 & \multicolumn{7}{c}{Stellar density [stars arcsec$^{-2}$]} 		\\
	    & 100      & 300      & 1000    & 3000    & 10000    & 30000    & 100000    \\
	\hline
	8 kpc    & 0.02\,M$_\odot$      & 0.03\,M$_\odot$     & 0.07 \,M$_\odot$   & 0.15\,M$_\odot$    & 0.5 \,M$_\odot$     & 1.1\,M$_\odot$      & 5        \,M$_\odot$ 	\\
	20 kpc   & 0.04\,M$_\odot$     & 0.05\,M$_\odot$     & 0.09\,M$_\odot$    & 0.17\,M$_\odot$    & 0.5\,M$_\odot$      & 1.1\,M$_\odot$      & 5\,M$_\odot$        	\\
	50 kpc   & 0.09\,M$_\odot$     & 0.11\,M$_\odot$     & 0.15\,M$_\odot$    & 0.2\,M$_\odot$     & 0.5\,M$_\odot$      & 1.1\,M$_\odot$      & 5\,M$_\odot$         	\\
	200 kpc  & 0.5\,M$_\odot$      & 0.5 \,M$_\odot$     & 0.5  \,M$_\odot$   & 0.5\,M$_\odot$     & 0.5 \,M$_\odot$     & 1.1\,M$_\odot$      & 6\,M$_\odot$         	\\
	800 kpc  & 1.1\,M$_\odot$      & 1.1\,M$_\odot$      & 1.1\,M$_\odot$     & 1.1\,M$_\odot$     & 1.1 \,M$_\odot$     & 2\,M$_\odot$        & 6\,M$_\odot$         	\\
	2 Mpc    & 2\,M$_\odot$        & 2\,M$_\odot$        & 2\,M$_\odot$       & 2\,M$_\odot$       & 2\,M$_\odot$        & 3\,M$_\odot$        & 7\,M$_\odot$         		\\
	5 Mpc    & 7 \,M$_\odot$       & 7\,M$_\odot$        & 7\,M$_\odot$       & 7\,M$_\odot$       & 7 \,M$_\odot$       & 7\,M$_\odot$        & 9\,M$_\odot$    			\\    
	\hline
	\end{tabular}
\end{table}

\begin{figure}
	\centering
	\includegraphics[width=0.7\textwidth]{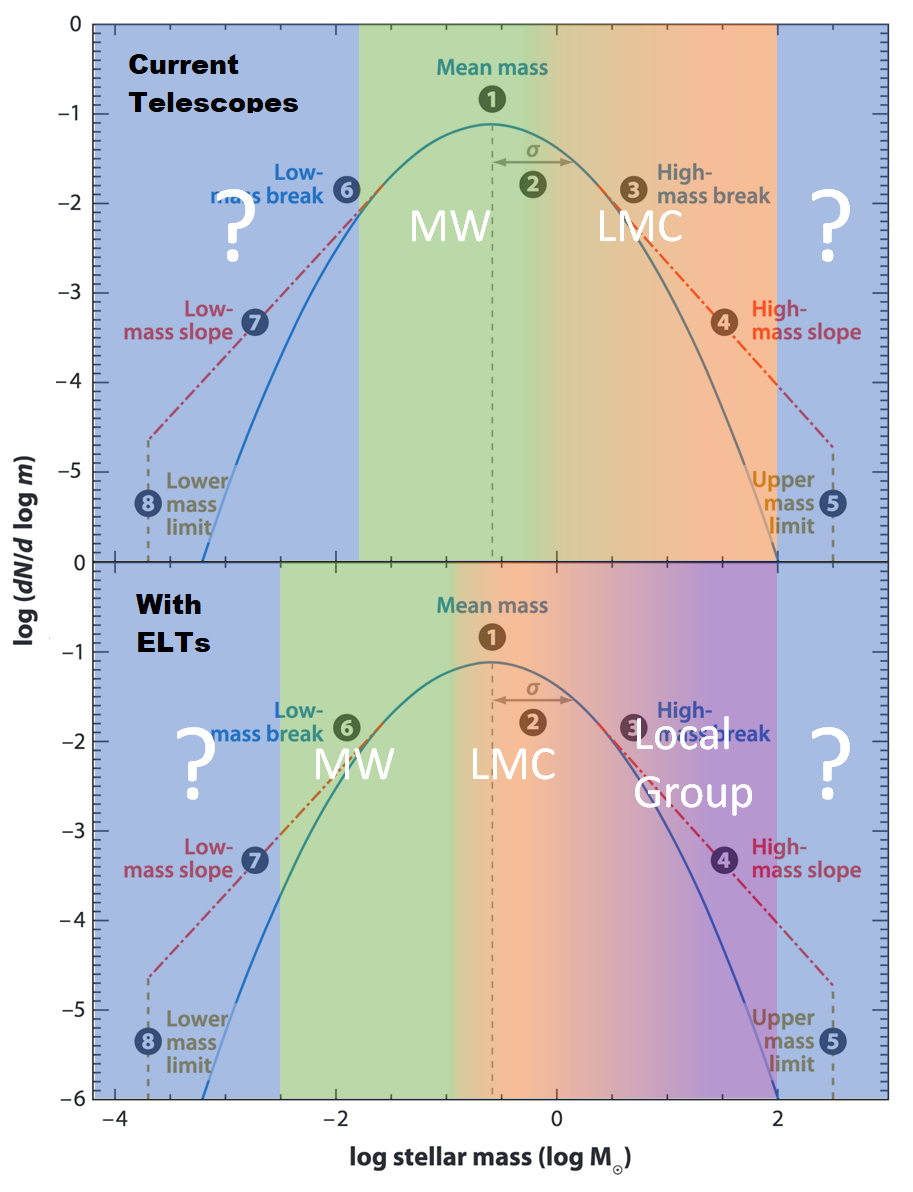} 
	\caption{The regions of the IMF at different distances which are accessible to the current generation of telescopes and which will be accessible to the future ELTs. While contemporary studies of the IMF are limited to the heavy side of the turnover around 0.4\,M$_\odot$ outside the milky way (e.g. in the LMC), the ELTs will have the resolution and sensitivity to study the shape of the low mass IMF in the LMC. They will also allow direct measurements of the high-mass slope in most galaxies within 5\,Mpc. The underlying image was taken from Bastian et al.(2010)}
	\label{fig:imf_comparison}
\end{figure}

\section{Conclusions}

The ELTs should enable us to answer two of the biggest open questions regarding the IMF: constancy and universality. Does the shape of IMF remain constant irrespective of environmental conditions? Is the shape of the IMF the same in all parts of the universe. Additionally the ELTs will enable observations that can bridge the two techniques currently used to determine the structure of the IMF inside and outside the Milky Way by being able to detect and resolve a sufficient number of individual stars in extra-galactic star forming regions out to 5\,Mpc.

Figure \ref{fig:imf_comparison} illustrates the improvement in observational capabilities regarding the IMF that will be brought about by the ELTs. Through this study we have shown that MICADO at the ELT will be capable of: 1. studying the brown dwarf mass function in almost any young cluster in the milky way, 2. observing in great detail the low mass IMF and the structure of the turn over around 0.4\,M$_\odot$ in our galactic neighbourhood, 3. opening up the densest regions in the cores of massive young clusters in the LMC and beyond, and 4. constraining the slope of the solar- and super-solar IMF in galaxies in and beyond the local group.

\end{document}